# High Harmonic Generation from Isolated Bound States: Tunable Emission Spectra and Mollow Triplets

Sergey Hazanov[1,2][†], Alexey Gorlach[1][†], Ron Ruimy[1], Dmitry Yakushevskiy[1], Marcelo F. Ciappina[1,3], and Ido Kaminer[1]

[1] *Technion – Israel Institute of Technology, Haifa 3200003, Israel*
[2] *Weizmann Institute of Science, Rehovot 7610001, Israel*
[3] *Guangdong Technion – Israel Institute of Technology, Shantou, Guangdong 515063, China*
Author e-mail address: kaminer@technion.ac.il

[†] *equal contribution*

**Abstract**

High harmonic generation (HHG) is a highly nonlinear emission process in which systems driven by intense laser pulses emit integer multiples (harmonics) of the driving field. This feature is considered universal to all occurrences of HHG. Here we show that a strong nonlinear response of certain systems can split the HHG spectrum into non-harmonic Mollow-type triplets, imprinting the internal electronic characteristics and confining potential on the HHG spectrum. The spectral lines become tunable by the driver field. These phenomena are universal to any material system with isolated bound states. We identify the conditions under which our predictions become accessible and propose experimental systems to observe them.

**Section I – Introduction**

When driven by an intense laser pulse, different types of systems such as atoms, moleculues, and more recently solids, can emit photons at frequencies that are integer multiples of the driving field. This highly nonlinear effect is called high harmonic generation (HHG) [1-3]. The electron dynamics during HHG is usually well-described by the three-step model [2]: An electron escapes the potential barrier of the atom, accelerates by the driving field, and recombines with the nucleus to emit harmonics.

The high harmonic spectrum has characteristic features: its frequency range extends far beyond the ionization potential of the source atom with fairly uniform intensity (the so-called plateau) and then abruptly falls off (the cutoff). The HHG spectrum contains only odd harmonics of the driving frequency $\omega_d$ due to the inversion symmetry of the atomic target [4],

$$\omega_{OHG}^{(n)} = (2n+1)\omega_d, \quad n \in \{1, 2, \ldots\}. \tag{1}$$

Still preserving this symmery, a global spectral shift was observed as a result of a change in the carrier-envelope phase [5]. Symmetry breaking leading to even harmonics generation was shown in theory [4,6-9] and experiments [10,11]. Interestingly, previous studies that considered double-well potentials and $H_2^+$ molecules predicted both even harmonics and non-harmonic peaks in the spectrum without any symmetry breaking [12-14]. To the best of our knowledge, none of these predictions have been observed so far.

The mathematical origin of some spectral features of HHG can be traced back to an elementary model: the two-level system (TLS). Driving a TLS with a time-harmonic potential creates a phenomenon akin to HHG [15-18] (see Sec. II in the SM). There, besides the odd harmonics $(2n+1)\omega_d$, one finds non-harmonic spectral lines [19-21]. But these predictions have never been observed. The lack of experiments made the non-integer harmonic features seem like artifacts of an oversimplified mathematical model. It remains unknown whether such features could have a physical origin; their fundamental nature and relevance to experiments call for further exploration.

Here we find the physical mechanism that can spawn non-harmonic features in HHG spectra: arising from isolated bound electron states. Our findings provide the recipe for future experiments that could observe such non-harmonic features. These features have a universal nature that can appear in a wide range of systems, connecting the math of the double-potential well and the TLS to realistic implementations in solid-

HHG, in HHG driven by THz pulses, and in HHG from ionized atoms and molecules [22-24]. Recent experimental advances in HHG in such materials open opportunities for the first observation of our predictions in readily available platforms.

Altogether, our work highlights a yet unobserved regime of HHG arising from nonlinear dynamics of bound electron states. This new regime has a direct signature in the HHG spectrum, in the form of Mollow triplets [25] that appear as a triple-splitting of each harmonic. Fig. 1 shows how these features are tunable by the driver intensity, imprintinh the shape of the potential and the structure of energy levels on the HHG spectrum. We identify the universal nature of this effect by demonstrating it in a range of potential wells. These singular HHG regimes can lead to applications in attoscience, X-ray physics, and coherent control of core electrons, unveiling new processes in strongly driven systems and providing a better understanding of existing ones.

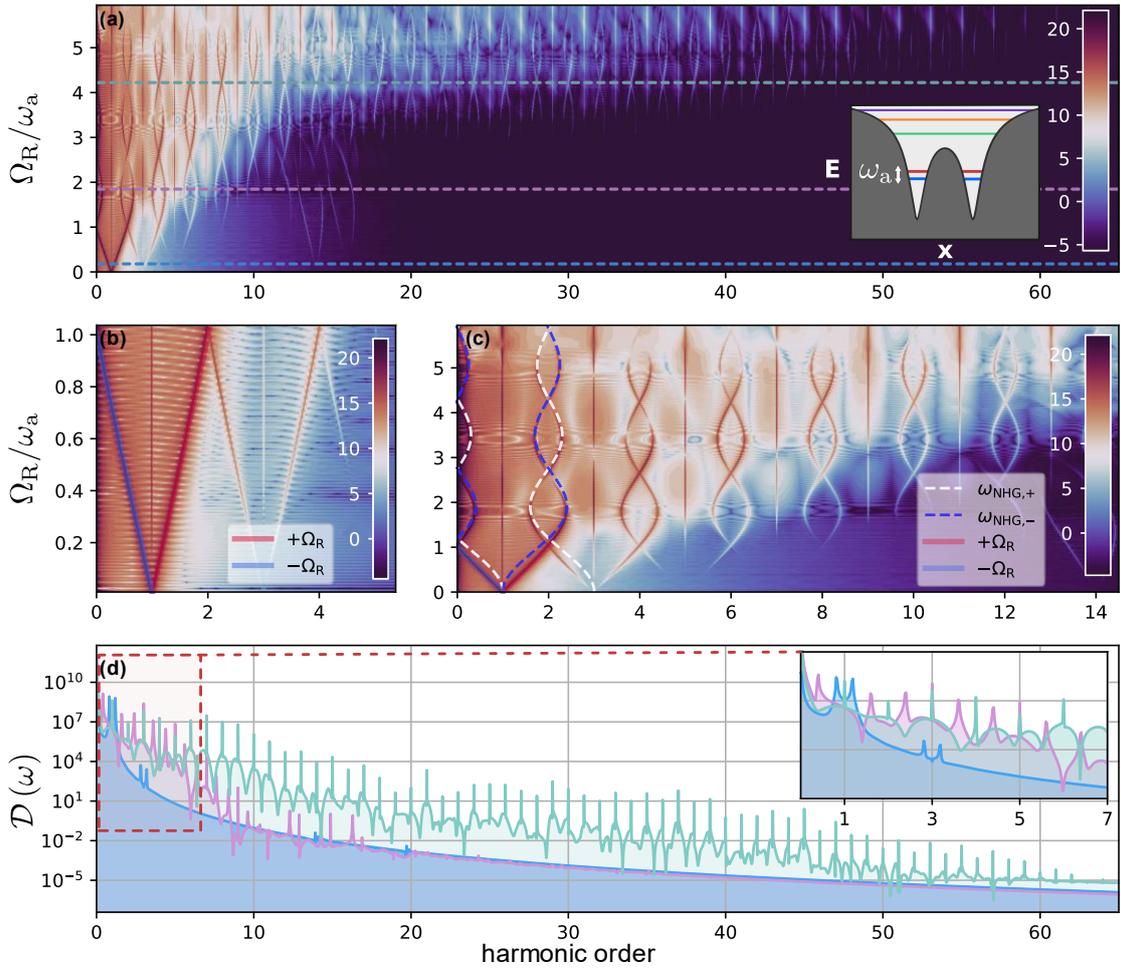

**Fig. 1 High harmonic generation (HHG) spectrum with non-harmonic features arising from the nonlinear response of isolated bound electron states.** (a) HHG spectrum of an electron in a double-Coulomb potential driven by a field with a frequency $\omega_d$ resonant with the frequency of the transition between the ground and the first excited state $\omega_a$ ($\omega_d = \omega_a$). The horizontal axis is the harmonic order (frequency divided by $\omega_d$). The vertical axis is the driving amplitude in terms of the Rabi frequency $\Omega_R$

of the two-level sub-system (denoted by blue and red lines in the inset) with energy separation $\hbar\omega_a$. The colormap indicates the dipole spectrum on a logarithmic scale, $\log \mathcal{D}(\omega)$ (see Sec. I in the SM). The three horizontal lines correspond to the spectra depicted in (d). **(b)** First and third-order harmonics in the nonlinear optics regime, dressed by the corresponding Mollow triplets, approximated by $(2n + 1)\omega_d \pm \Omega_R$ for $\Omega_R < \omega_a$ and $n = 0,1$. **(c)** Low-order harmonics in the extreme nonlinear regime. The helix-shaped spectral lines are the carrier-wave Mollow triplets, which for $\Omega_R > \omega_a$ are well approximated by Eq. (2). **(d)** Three emission spectra: below the extreme nonlinear regime (blue) and in the extreme nonlinear regime (purple and green). For $\Omega_R < \omega_a$, the first and third harmonics exhibit two sidebands with separation growing linearly with $\Omega_R$. In the extreme nonlinear regime, the sidebands' frequency is determined by Eq. (2). It can be purely harmonic, with the sidebands from adjacent odd-harmonics coinciding with the even-harmonic (green) or non-harmonic (purple). The maximal field amplitude (in (a) and (c)) corresponds to $\Omega_R \approx 6\omega_a$ or an intensity of $I_0 = 1 \times 10^{14}$ W/cm².

**Section II – Dressing the spectrum of HHG**

To motivate the study of HHG from non-Coulomb-like potentials, we begin by revisiting the nonlinear response of a TLS in an intense external drive. Consider a TLS of energy separation $\hbar\omega_a$, driving field frequency $\omega_d$, and a corresponding Rabi frequency $\Omega_R$. Studying the emission spectrum of this system without the rotating-wave approximation, one finds a significant change in the system's response when $\Omega_R$ is brought close to the energy separation $\omega_a$ and exceeds it [18]. In the case of a resonant drive with $\Omega_R \lesssim \omega_a$, we identify a single dominant Mollow triplet [26], whose central spectral line $\omega = \omega_d$ is dressed by two sidebands at frequencies $\omega = \omega_d \pm \Omega_R$.

In the extreme nonlinear regime, the Rabi frequency surpasses the atom's energy separation $\Omega_R > \omega_a$ and the emission spectrum admits additional spectral triplets (which in the resonant case are known as carrier-wave Mollow triplets [18, 27]). Then, the central lines coincide with the odd harmonics of the driving frequency and are independent of the Rabi-frequency. In this regime, the two sidebands appear at non-harmonic frequencies and within the Floquet theory, these frequencies are approximated by [13]

$$\omega_{\text{NHG},\pm}^{(n)} = (2n + 1)\omega_d \mp \left[\omega_d - \omega_a J_0\left(2\Omega_R/\omega_d\right)\right], \qquad (2)$$

where $n \in \{1, 2, ...\}$ and $J_0$ is the zeroth-order Bessel function of the first kind. Note that for small $\omega_a$, the sidebands appear to surround the even harmonics.

Below we show how the Mollow-like sidebands can be imprinted on the spectrum of HHG in a realistic physical system. To realize this effect, we consider an atomic system that has two (or more) isolated bound states, having energy separation much smaller than the ionization energy, $\omega_a \ll E_{\text{ion}}$. This condition imitates the ideal TLS by isolating the bound states from the ionization threshold. Crucially, this condition excludes Hydrogen atom-like potentials in which the energy gap between the ground and the first excited state (10.2 eV) is of the same order as the ionization energy (13.6 eV). Natural candidates for satisfying the above conditions are multi-well potentials (Fig. 1), in which the ground and first excited states are close to each other.

Fig. 1 depicts the emission spectrum of a double-well potential driven by an intense resonant field. Figs. 1(b-c) show that the features predicted in a TLS can describe this system: for driving field satisfying $\Omega_R < \omega_a$, the sidebands' frequency grows linearly with the field amplitude; for $\Omega_R > \omega_a$, the non-harmonic frequency is given by Eq. (2) (Sec. III of the SM). Figure 1(d) depicts the emission spectrum of three different Rabi frequencies and shows that within this HHG regime, the emission spectrum may admit both harmonic and non-harmonic spectral lines.

The emission spectrum depicted in Fig. 1 is yet to be observed experimentally. This is because the ubiquitous Coulomb potential does not support isolated electronic states that have a strongly nonlinear response without getting ionized. Figure 2(a-c) compares the emission spectra of a single-, a double- and a triple-Coulomb well potential under a non-resonant intense external field. In Fig. 2(a), an electron bounded by a single-well Coulomb potential and driven by a laser with $\omega_d = 1.55$ eV is shown to emit only odd harmonics of $\omega_d$. On the contrary, for an electron in the double-well potential driven by a field with $\omega_d = 0.7\omega_a$ (see Fig. 2(b)), the non-harmonic generation is dominant, having intensity comparable with the odd-harmonics. In the extreme nonlinear regime, the non-harmonic spectral lines are in good agreement with Eq. (2).

The double-well potential is the simplest case suitable for non-harmonic generation. More complicated well-type structures give rise to more intricate spectra. We consider a triple-well potential such that the three lowest eigenenergies are equidistant, $\Delta \mathcal{E}_{10} \approx \Delta \mathcal{E}_{21} \equiv \hbar \omega_a$. Driving this system with a pulse of $\omega_d = 0.7\omega_a$ results in the emission spectrum depicted in Fig. 2(c). This spectrum exhibits contributions from three individual TLSs: ground and first-excited state, first excited state and second-excited state, and finally the ground and second excited state having energy separation $2\omega_a$. The light-blue and red dashed lines in Fig. 2(c) correspond to Eq. (2) evaluated for $\hbar \omega_a = \Delta \mathcal{E}_{10} \approx \Delta \mathcal{E}_{21}$. Interestingly, the spectral lines of the TLS with frequency $2\omega_a$ intersect the odd-harmonics and cannot be fitted by Eq. (2). These spectral lines are found to satisfy a modified version of Eq. (2) in which the Mollow triplets appear around the odd harmonics rather than the even ones:

$$\widetilde{\omega}_{\text{NHG},\pm}^{(n)} = (2n+1)\omega_d \pm \omega_a^{(3)} J_0\left(2\Omega_R/\omega_d\right), \qquad (3)$$

where $\omega_a^{(3)} \approx 2\omega_a$ (see Sec III for the general case) $n \in \{0, 1, 3 \ldots\}$.

The contributions arising from the two equidistant energy levels are predicted by Eq. (2) to be degenerate. Instead, Fig. 2(c) shows that the degeneracy is lifted by a small 'repulsion' between the energy levels. This may be analogous to the effect of avoided crossing in many physical systems [28]. This analysis can be taken beyond individual TLSs, to a many-level system, and eventually to a continuum of energy levels. There, each spectral line is repelled by the contributions from TLSs with similar energy separation, leading to an overall vanishing emission. This is why the Mollow triplet features gradually vanish when the number and the density of levels increase – eventually reproducing the conventional spectrum of HHG with only odd harmonics.

We emphasize that the non-harmonic spectral features arising from the isolated bound states allow engineering the emission spectrum of a driven system via manipulation of the energy levels structure and the driving intensity. This fact differs from the conventional odd-harmonics that are uninfluenced by the shape of the potential.

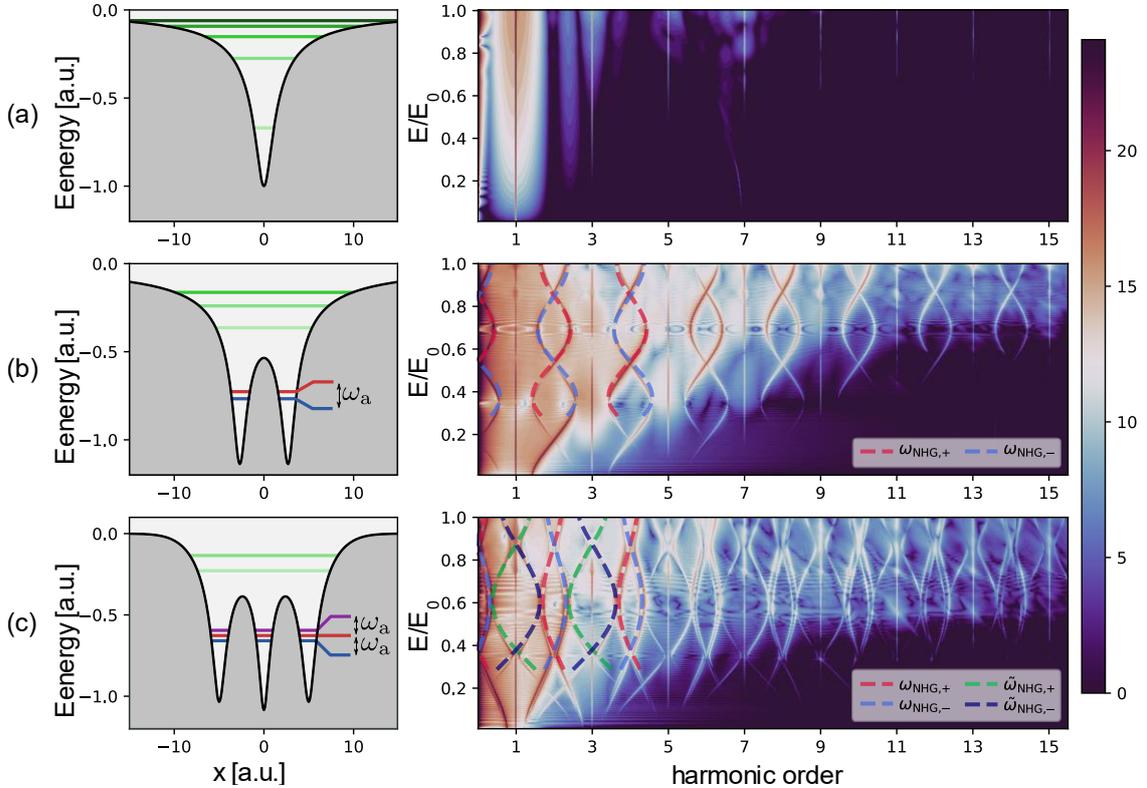

**Fig. 2 Tunable non-harmonic spectral lines.** The color map indicates the dipole spectrum on a logarithmic scale, $\log \mathcal{D}(\omega)$ (Sec. I in the SM). **(a)** Single-Coulomb potential (left) and the corresponding traditional HHG spectrum, admitting only odd harmonics. **(b)** Double-Coulomb potential: the conventional HHG spectrum of odd harmonics is supplemented by the non-harmonic lines caused by the extreme nonlinear response of the ground and first excited states, with energy separation $\omega_a$. **(c)** Triple-Coulomb potential: the spectrum has additional spectral lines described by Eqs. (2-3). The external pulse of trapezoidal shape is characterized by $(n_{\text{on}}, n_{\text{p}}, n_{\text{off}}) = (1, 50, 1)$ (Sec. I in the SM). In (a), the carrier frequency is $\omega_d = 0.057$ a. u. (1.55 eV) with $\Delta\mathcal{E}_{10} = 0.395$ a. u. (10.75 eV) and in (b-c), the carrier frequency satisfies $\omega_d = 0.7\omega_a$, with $\omega_a = 0.04$ a. u. (1.09 eV) in (b) and $\omega_a = 0.032$ a. u. (0.87 eV) in (c). Note that a. u. refers to atomic units here and in the other figures.

**Section III – Universality and experimental feasibility**

Previously we considered potentials tailored for reproducing the behavior of a strongly driven TLS, such as in Fig. 2(b). The external driving fields were chosen with a large number of driving cycles, amplifying the non-harmonic generation. Under these conditions, the non-harmonic spectral lines are strong and even sometimes dominant. Here we simulate more realistic scenarios, including a drive with a Gaussian profile and bound states with different energy gaps. Furthermore, we propose systems in which the discussed effect can be feasibly observed.

Figure 3(1a) depicts the emission spectrum from a double-well potential (Fig. 2(b)), driven by a Gaussian pulse with $\tau_{\text{FWHM}}$ occupying 20 optical cycles (Sec. I in the SM). For such a short pulse, the Rabi frequency cannot be approximated to be constant. In turn, this has the effect of averaging the helix-shaped patterns depicted in Fig. (2b). While the non-harmonic spectral lines are partially smeared, the TLS contribution to the spectrum can still be resolved, with the features being sharper for wider pulses.

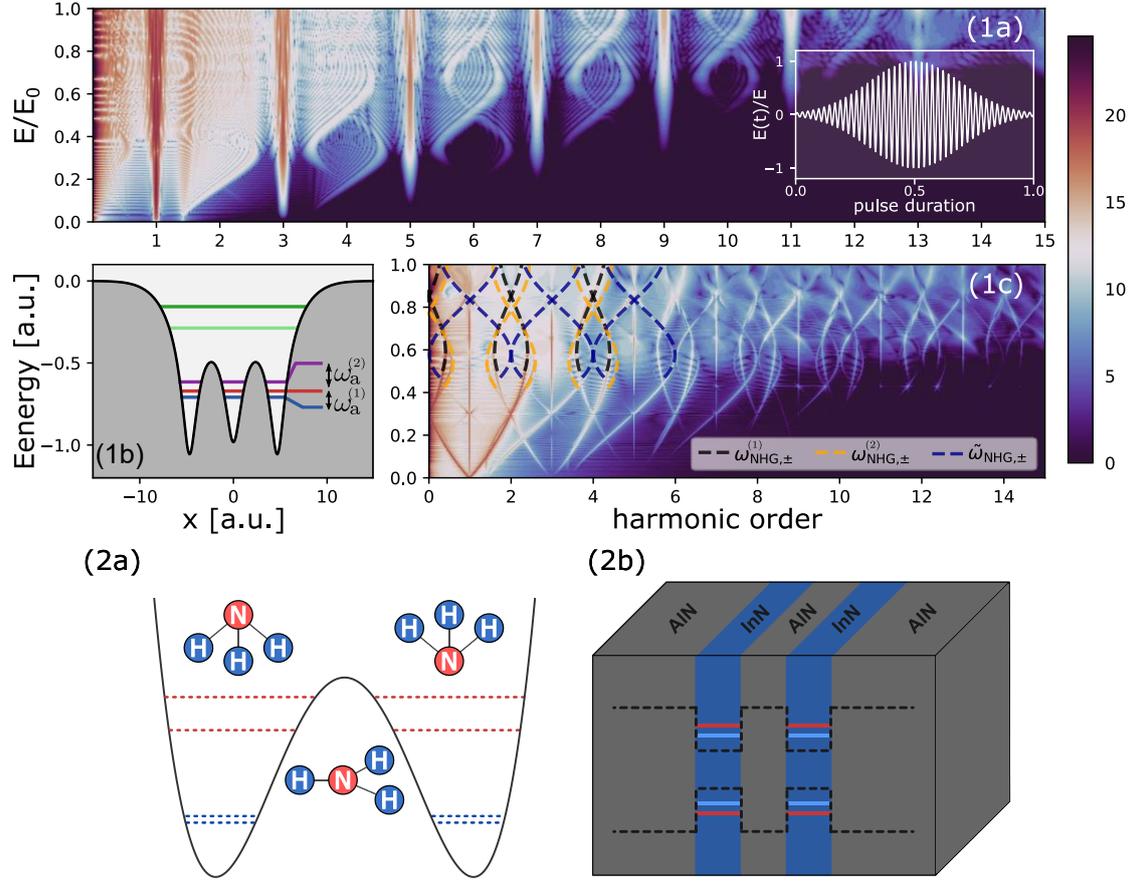

**Fig. 3 Experimental feasibility. (1a)** Electron in a double-well potential driven by a Gaussian pulse with $\tau_{\text{FWHM}} = 20$ optical cycles (shown in inset). The helix-shaped non-harmonic generation is smeared but its overall shape can still be resolved. **(1b-c)** Triple-well Coulomb potential with three isolated levels, with three different energy separations: $\omega_a^{(1)}, \omega_a^{(2)}$ and $\omega_a^{(1)} + \omega_a^{(2)}$. The contributions to the emission spectrum of the three sub-systems can be resolved by employing Eqs. (2-3). In (1a), the carrier frequency is $\omega_d = 0.7\omega_a$ with $\omega_a = 0.038$ a. u. (1.04 eV) and in (1b) $\omega_d = \omega_a^{(1)}$ with $\omega_a^{(1)} = 0.038$ a. u. (1 eV) and $\omega_a^{(2)} = 0.055$ a. u. (1.5 eV). **(2a)** Ammonia molecule rotational spectrum, containing isolated energy levels with MW (blue) and IR (red) frequency transitions. **(2b)** Semiconductor heterostructure comprised of 3-layers of aluminum-nitride and 2-layers of indium-nitride, reconstructing the double-well structure.

Figure 3(1b-c) depicts an atomic system in which the energies of the three lowest bound states ($\mathcal{E}_0, \mathcal{E}_1, \mathcal{E}_2$) are not equidistant. In this case, the contributions from the different bound state transitions will result in spectral lines having substantially different frequencies. These spectral lines can be explained by constructing three TLSs $\hbar\omega_a^{(1)} \equiv \mathcal{E}_1 - \mathcal{E}_0$, $\hbar\omega_a^{(2)} \equiv \mathcal{E}_2 - \mathcal{E}_1$ and $\hbar\omega_a^{(3)} \equiv \mathcal{E}_2 - \mathcal{E}_0$ and employing Eq. (2) for the first and second TLSs and Eq. (3) for the third TLS.

To observe this novel regime of HHG, one needs to consider systems admitting bound states isolated from the ionization threshold. The Coulomb potential is an inadequate candidate for this purpose because the energy separation between its ground and first excited states is of the same order of magnitude as the ionization energy, which implies quick ionization under a strong drive. One possible route to overcome this limitation is HHG from molecules. For example, the inversion-rotation-vibration energy levels of the Ammonia molecule, depicted in Fig. 3 (2a), have characteristics similar to the double-well potential discussed previously. In addition, these isolated eigenenergies are also tunable by a DC electric field [29], allowing to simplify the experimental effort to drive the corresponding TLSs. Another example is the $H_2^+$ ion, which was previously studied in the context of other non-harmonic spectral features [13, 30]. Finally, a promising platform for demonstrating this regime is quantum wells constructed from semiconductor heterojunctions [31, 32]. In such systems, the energy levels can be engineered by choosing appropriate materials and tuning the width of the semiconductor layers (e.g., see the AlN-In structure depicted in Fig. 3(2b)).

**Section IV – Inter-potential dynamics**

The previous sections showed that the shape of the stationary potential affects the system's nonlinear response, leading to an intricate emission spectrum. This emission can be understood in terms of multiple TLS responses of the bound states. We now show how this response can be explained by the evolution of the electron's wavepacket. Figure 4(a-b) depicts the evolution of the electron's wavepacket in a Coulomb potential, where the driving field creates HHG but simultaneously depletes the atom. The depletion occurs before any significant contribution of non-harmonic generation.

Figure 4(c-d) depicts multi-well stationary potentials. In these potentials, driving amplitudes that do not cause depletion can still create HHG. The prolonged non-depleting dynamics cause the electron dynamics to undergo sub-optical-cycle transitions between the wells, leading to the emission of non-harmonic sidebands. We conjecture that given more complicated stationary potentials, comprised of more isolated bound states and different well shapes, the emission spectrum may take even more elaborated forms.

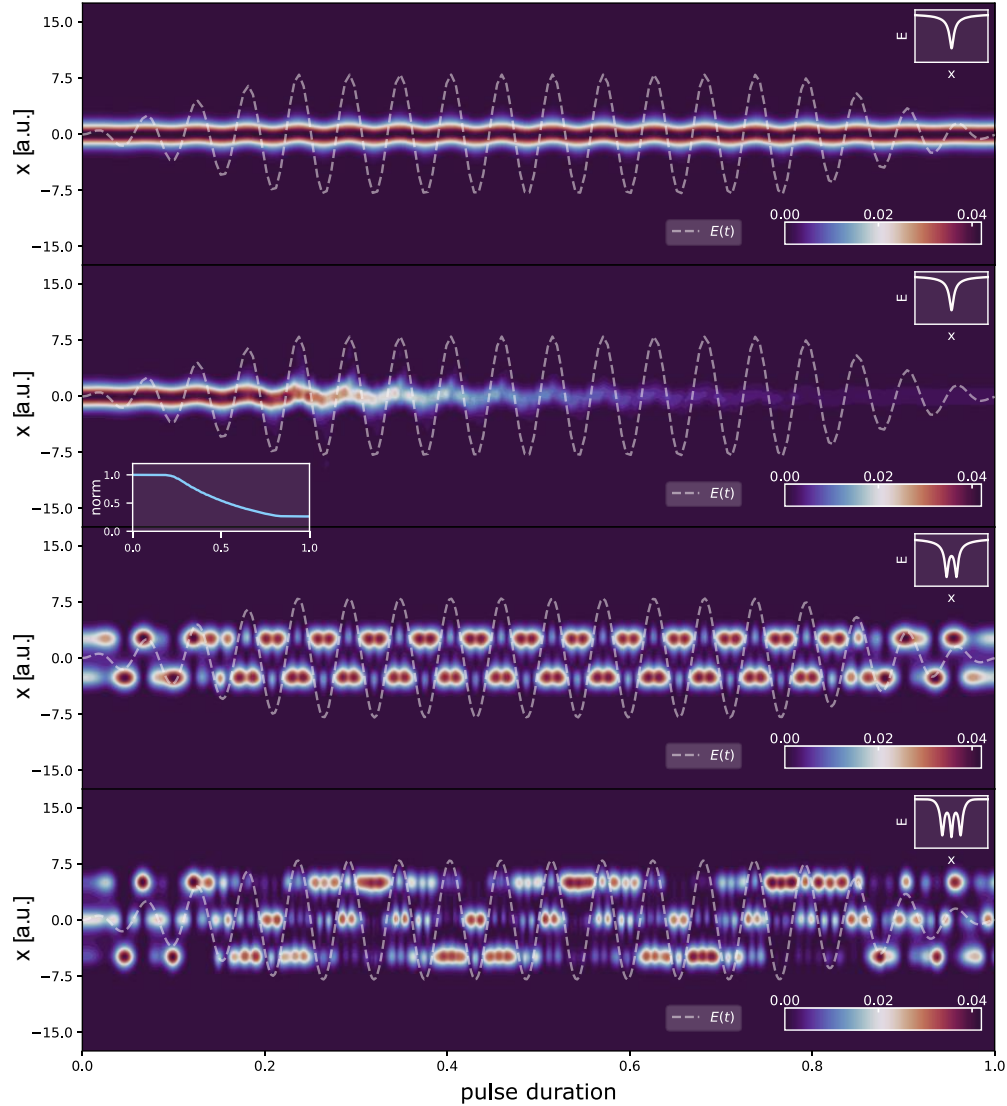

**Fig. 4 Inter-potential wavepacket dynamics as the mechanism of non-harmonic spectral lines. (a-b)** Time evolution of the probability density $|\Psi(x,t)|^2$ in the Coulomb potential (Fig. 2(a)). Driving the system with $E = E_0$ results in mild tilting of the electron's position, whereas with $E = 1.8E_0$, ionization becomes important, gradually depleting the wave function to the continuum (as depicted in the left inset), simultaneously with the generation of HHG. The dashed line represents the phase of the driving field. **(c-d)** Time evolution of double- and triple-well potentials, with the driving frequency far below the ionization frequency, showing strong oscillations of $|\Psi(x,t)|^2$ without depletion due to ionization. These rapid, non-depleting oscillations result in the intricate non-harmonic spectral lines. The corresponding potential wells are marked by the right insets. In (c-d), we simulate the same systems as in Fig. 2(b-c) with $E = 0.5E_0$.

## Section V – Conclusion

We conclude that an essential condition for non-harmonic peaks generation is the ability of an atomic potential to support bound states isolated from the ionization threshold. This condition enables to drive two (or more) of the potential's bound states for a long enough duration without complete depletion to ionization. This prerequisite on the shape of the binding potential lays down restrictions on the set of realistic scenarios in which this phenomenon can be observed. Notably, the Coulomb potential does not support non-harmonic peaks generation, while multi-well potentials can be used for tunable non-harmonic generation. We showed that the emission of amplitude-dependent non-harmonic frequency peaks, hitherto mainly attributed to the ideal TLS and the double-well potential, is common for many generic scenarios. The intricate HHG spectra in such cases can be understood either using one or more TLS models or by considering the electron's wavepacket time evolution.

The non-harmonic spectral lines we found are a simple form of a broader phenomenon, foretelling an ample variety of non-harmonic spectra. These highly elaborated spectra can be achieved by considering systems with a larger number of isolated bound states. Looking forward, the rich physics of strongly-driven TLS can inspire other phenomena that may appear in HHG experiments, such as carrier-wave Rabi floppings [33]. Such effects are likely to be observed in systems that allow to engineer their bound states and can be potentially implemented in molecular and solid-state systems. We believe that non-harmonic generation tunable via the driving field's adjustment is a valuable tool in attosecond science, promising tunability of X-ray sources, quantum coherent control of core electrons, and generation of attosecond pulses with desired properties.

## Supplementary material
### Section I - Theoretical model and simulation framework

Throughout this work, we employ a 1D model in which an electron is bound to a stationary potential and driven by an external, linearly polarized, classical field. The time-dependent Schrodinger equation (TDSE) is integrated via the Crank-Nicolson method [1]. In addition, to avoid reflections from the boundaries, the wave function is multiplied by a mask function which decays as $\sin^{18}(x) \in [\pi/2, \pi]$ at the boundaries [2]. This mechanism also introduces a decrease in wave function's normalization, effectively modeling depletion of the driven atom.

The high-harmonic generation (HHG) spectrum is evaluated within the single-active electron approximation [3]. The atomic part of the Hamiltonian is given by

$$H_a(x) = \frac{\hbar^2}{2m} \frac{d^2}{dx^2} + \mathcal{V}_a(x), \tag{1}$$

where $m$ is the mass of the electron and $\mathcal{V}_a(x)$ is the stationary potential to which the electron is confined. The eigenenergies pertaining to the isolated bound states (the two eigenenergies contributing to the non-integer harmonics) are denoted by $\mathcal{E}_g$ and $\mathcal{E}_e$, and the corresponding energy separation is given, in frequency units, by $\hbar\omega_a = |\mathcal{E}_e - \mathcal{E}_g|$.

The external driving field is introduced by adding the following interaction term to the Hamiltonian

$$H_I(x, t) = -exE(t), \tag{2}$$

where $E(t) = p(t)E\sin(\omega_d t)$, $p(t)$ is the pulse envelope, $E$ is the maximal amplitude of the pulse and $\omega_d$ is the driving frequency. We consider both a Gaussian envelope, given by

$$p(t) = \exp\left[-\frac{4\log 2}{\tau_{FWHM}^2}\left(t - \frac{\tau}{2}\right)^2\right], \tag{3}$$

where $\tau_{FWHM}$ is the FWHM duration, and a trapezoidal envelope, given by

$$p(t) = \begin{cases} \dfrac{t}{\tau_{on}}, & 0 < t < t_{on} \\ 1, & t_{on} < t < \tau - \tau_{off}, \\ -\dfrac{t-\tau}{\tau_{off}}, & \tau - \tau_{off} < t < \tau \end{cases} \tag{4}$$

where $\tau$ is the duration of the entire pulse and $\tau_{on}$ ($\tau_{off}$) is the turn on (off) period of the pulse. The above durations are defined by the number of periods of the driving field, $\sin(\omega_d t)$, occupying each duration. Therefore, a trapezoidal pulse envelope is fully characterized by the tuple $(n_{\text{on}}, n_{\text{p}}, n_{\text{off}})$, where $n_I$ is the number of periods in the corresponding duration. Similarly, the pulse envelope of a Gaussian pulse is determined by $n_{\text{FWHM}}$. The relation between $I$ and $\tau_I$ is given by $\tau_i = 2\pi n_i/\omega_d$.

The total time-dependent Hamiltonian reads

$$H(x,t) = H_a(x) + H_I(x,t). \tag{5}$$

Employing the dipole acceleration, which is given by

$$\ddot{d}(t) = \frac{d^2\langle ex \rangle}{dt^2} = -e\langle \Psi(x,t)|[H(x,t),[H(x,t),x]]|\Psi(x,t)\rangle, \tag{6}$$

the emission spectrum is proportional to [4,5]

$$\mathcal{D}(\omega) = \left|\frac{1}{\omega^2}\int_{-\infty}^{\infty} dt\, \ddot{d}(t)e^{-i\omega t}\right|^2. \tag{7}$$

The stationary potentials employed in the previous sections, $\mathcal{V}_a(x)$, are comprised of combinations of single soft-Coulomb potential wells, separated from each other by distance $d$. Each such well is given by

$$\mathcal{V}_{\text{s.c}}(x) = \frac{1}{\sqrt{ax^2+b}}, \tag{8}$$

where $a$ and $b$, along with $d$, are parameters that determine the bound states of the entire stationary Hamiltonian.

Controlling the number of potential wells and the values of $a$ and $b$ affects the eigenenergies of the isolated bound states and allows to fine-tune the non-harmonic spectral lines of the emitted radiation. This feature can be further enhanced by exploiting systems where the characteristics of the potential can be altered externally (such as an additional electromagnetic field, as in the Ammonia molecule). Stacking this ability on top of the spectrum's sensitivity to the intensity of the driver field, this phenomenon results in additional degrees of freedom to tune and control the source of radiation.

**Section II – The rotating wave approximation and the two-level system model**

Typically, the rotating wave approximation (RWA) and the two-level system (TLS) model go hand in hand in atomic systems, as stated by [6]: "Something that is sometimes not appreciated is that the two-level approximation and the RWA are at the same level of accuracy. That is, it makes no sense to throw one out and keep the other. Both amount to discarding interactions that are far off-resonance (the RWA amounts to a very far detuned interaction of a positive-frequency resonance with a negative-frequency field). If the detuning is large enough that the counter-rotating term is not negligible, then neither are the couplings to the other levels."

The usual conclusion is that the non-harmonic features in the TLS model are an artifact of an oversimplified model. Our work does not contradict this conclusion. Instead, we search for physical conditions under which the signature features of the TLS would appear in realistic experiments. Our conclusion is that isolated bound states could create such features, appearing as the non-Harmonic spectral lines called Mollow Triplets.

**Section III – Dipole moment in a 1D system**

To estimate the dipole moment pertaining to an isolated TLS in a 1D potential well, we fit a linear curve to the linear part of the sidebands' spectrum, where the Rabi frequency is smaller than the energy separation of the TLS, $\Omega_R < \omega_a$. The sidebands frequencies are given by $\omega_d \pm \Omega_R$ where $\Omega_R = d \cdot E$. The field amplitude is approximated to be constant, seen here using a trapezoidal envelope with a short rise and fall durations. For example, in Fig. (1) in the main text, the pulse's parameters are given by $(n_{on}, n_p, n_{off}) = (1, 50, 1)$ (see Sec. I).

In the above formula, the value of the dipole moment $d$ is fitted according to the linear part of the non-harmonic spectrum, and the field's amplitude is approximated to be constant. This is a reasonable approximation for long pulses (typically, tens of femtoseconds and above). We specifically use a trapezoidal field envelope with $(n_{on}, n_p, n_{off}) = (1, 50, 1)$ (see Sec. V).